\documentclass{PoS}

\pdfoutput=1

\newcommand{\pbp}{\langle\bar\psi\psi\rangle}

\title{Lattice QCD with Eight Degenerate Quark Flavors}

\ShortTitle{Lattice QCD with Eight Degenerate Quark Flavors}

\author{\speaker{Xiao-Yong Jin}\\
        Department of Physics, Columbia University, New York, NY 10027, USA\\
        E-mail: \email{xj2106@columbia.edu}}

\author{Robert D. Mawhinney\\
        Department of Physics, Columbia University, New York, NY 10027, USA\\
        E-mail: \email{rdm@physics.columbia.edu}}

\abstract{We report on simulations of QCD with many flavors of
  degenerate quarks, the DBW2 gauge action and naive staggered
  fermions, using the rational hybrid Monte Carlo algorithm.  We
  primarily focus on eight degenerate quark flavors where a
  variety of values of the coupling constant and quark mass have
  been used in the simulations.  The scaling behavior of the
  hadron spectrum and the string tension of the heavy quark
  potential is studied, to probe whether the zero temperature,
  continuum limit of the theory breaks chiral symmetry.}

\FullConference{The XXVI International Symposium on Lattice Field Theory\\
  July 14 - 19, 2008\\
  Williamsburg, Virginia, USA}

\begin{document}

\section{Introduction}
\label{sec:introduction}

It is well known that vector-like gauge theories lose asymptotic
freedom if the number of massless fermion species, $N_f$, exceeds
some critical value $N_f^*$, where $N_f^* = 16.5$ for QCD.  For
$N_f$ large, but smaller than $N_f^*$, a weak coupling, infrared
fixed point exists in the two-loop beta function \cite{Caswell:1974gg,
Banks:1981nn}, which indicates that as $N_f$ is increased from zero
an interesting conformal phase may exist \cite{Dietrich:2006cm}
before $N_f$ becomes large enough to lose asymptotic freedom.  The
range of $N_f$ where such a conformal phase may exist is referred
to as the conformal window, and lattice simulations provide an ideal
method to determine the location of this window.  Using the
Schr\"odinger functional method, the running of the coupling constant
can be measured on the lattice \cite{Appelquist:2007hu}, which has
given evidence for a conformal window for QCD for $12 \leq N_f \leq
16$.  Additional studies of the behavior of the finite temperature
phase transition as a function of  $N_f$ can be used as a probe
into the conformal window \cite{Deuzeman:2008sc}.  In this paper,
we investigate the question of whether $N_f = 8$ QCD is in the
conventional, chirally broken phase of QCD, or in the conformal
phase.  Previous simulations of eight flavor QCD \cite{Brown:1992fz}
revealed lattice artifacts for the lattice spacings accessible at
that time and prevented a clear statement about the phase of the
theory at a single lattice spacing, and no information about the
continuum limit.  Here we use conventional, zero temperature lattice
simulations to measure the chiral condensate and light hadron
spectrum for a variety of quark masses and lattice spacings, so we
can study the zero quark mass limit and the continuum limit.

\section{Simulations and results}
\label{sec:simulations-results}

\subsection{Choice of algorithm: RHMC}
\label{sec:choose-algorithm}

We chose the Rational Hybrid Monte Carlo (RHMC) algorithm
\cite{Clark:2003na, Clark:2006fx}, because it is exact within machine
precision, its software implementation includes many optimizations.
and it is extensively used for our other $2+1$ flavor simulations
with domain wall and staggered fermions.  To check it for our current
task, we compared our results from a 4 flavor simulation, done as
a 2+2 flavor RHMC simulation (which requires the square root of the
staggered fermion determinant) to previous results with $\Phi$
algorithm \cite{Sui:2001rf}.  In Tab.~\ref{tab:rhmc_phi}, the
plaquette, $\pbp$, and masses of 4 meson channels are compared.\footnote{In
this paper, all numeric values of dimensionful quantities are in
lattice unit unless explicitly shown.}  The plaquette values agree
at the 1 $\sigma$ level, while the masses differ by 2 to 3 standard
deviations, making it likely that there are long autocorrelation
times in the simulations which are not under good control.

\begin{table}
  \centering
  \begin{tabular}{c|ll}
    Algorithm       &  $\Phi$        & RHMC           \\
    \hline\hline
    Run Length      & $1000\sim4760$ & $1290\sim3840$ \\
    Acceptance Rate & 0.87           & 0.60           \\
    Measurement Interval & 5         & 10             \\
    \hline
    Plaquette       & 0.560130(14)   & 0.560072(39)   \\
    $\pbp$          & 0.0404(1)      &  0.04105(35)   \\
    $m_\pi$         & 0.3210(40)     & 0.3127(30)     \\
    $m_{\pi_2}$     & 0.3543(35)     & 0.3495(41)     \\
    $m_\rho$        & 0.4763(59)     & 0.4610(31)     \\
    $m_{\rho_2}$    & 0.4777(84)     & 0.4561(37)     \\
  \end{tabular}
  \caption{A comparison between RHMC and $\Phi$ algorithm results using
  naive staggered fermions and the Wilson gauge action for $N_f=4$,
  $m_q=0.015$, $\beta=5.4$.}
  \label{tab:rhmc_phi}
\end{table}

\subsection{Choice of lattice action: Staggered fermions with DBW2 gauge}
\label{sec:choose-latt-action}

In this exploratory work, we seek to understand the basic properties
of $N_f=8$ QCD at zero temperature, in the chiral and continuum limits,
which requires simulations at many different parameter values.
We have chosen to use the na\"ive staggered fermion action for its remnant
chiral symmetry and its simulation speed.  To help control the
finite lattice spacing artifacts of staggered fermions, we have used
the DBW2 gauge action, which produces smoother gauge fields at the
lattice scale, for a given low energy physical scale, than other gauge
actions.  This smoothing of the gauge field might be expected to decrease
the lattice artifacts, which we will address shortly.

Additionally, using the DBW2 gauge action should help to compensate
for the coarsening of the gauge fields that comes from adding more
fermions.  From asymptotic freedom, we know that adding more fermions
to a lattice theory with a fixed low energy scale produces a larger
coupling at the lattice cutoff, and hence coarser gauge fields.
These coarse fields, in turn, make any lattice fermion formulation
a poorer approximation to the continuum.

\subsection{Measuring taste symmetry breaking}
\label{sec:taste-symm}

\begin{figure}
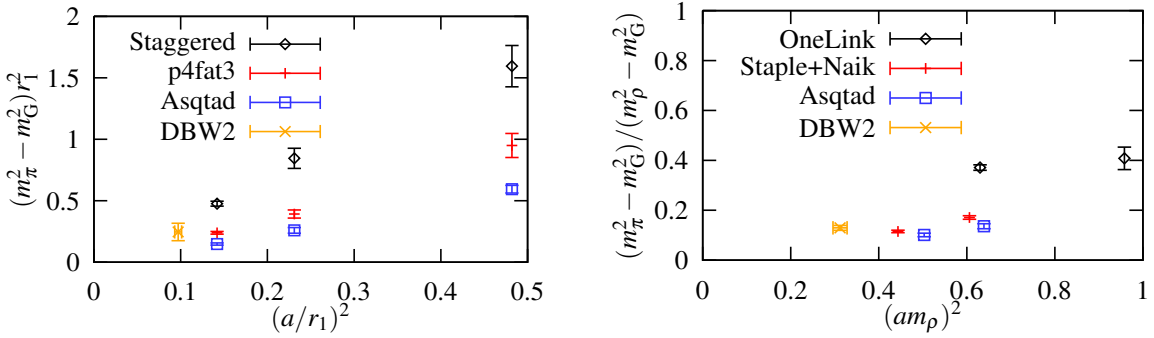

  \centering
  \includegraphics[width=0.47\textwidth]{fig/splitting_cheng.mps}
  \qquad
  \includegraphics[width=0.47\textwidth]{fig/splitting_milc.mps}
  \caption{A comparison of taste symmetry breaking.  Left: compare $N_f = 2$
  DBW2 results with quenched results from \cite{Cheng:2006wj}, where the
  chiral limit has been taken.  Right:
    compare $N_f = 2$ DBW2 results with $N_f = 2$ results from
    \cite{Orginos:1998ue,
      Orginos:1999cr} at $m_\pi/m_\rho = 0.55$.  Here, except for the 
      ``DBW2'' points, all labels refer to the type of valence quark used
      on a fixed dynamical ensemble.}
  \label{fig:split-cmpr}
\end{figure}

We have measured the taste symmetry breaking between the Goldstone
pion and the local, non-Goldstone pion for our DBW2 simulations
with na\"ive staggered fermions and 2 dynamical flavors.  In
Fig.~\ref{fig:split-cmpr}, our results are compared with results
from improved staggered actions in quenched \cite{Cheng:2006wj} and
2 flavor (``Staple+Naik'' sea quarks with Symanzik improved gauge
action) \cite{Orginos:1998ue, Orginos:1999cr} simulations.  Although
the comparisons are at different lattice spacings, it appears that
the DBW2 gauge action has reduced the splittings seen with
naive staggered fermions and the Wilson gauge action.

\subsection{Results with eight flavors}
\label{sec:eight-flav-res}

We report on simulations of QCD with 8 degenerate quark flavors
with three different values of the coupling: $\beta=0.58$, $\beta=0.56$,
$\beta=0.54$; 2 different lattice sizes: $16^3\times32$ and
$24^3\times32$; and 2 or 3 different values of the quark mass for
each coupling.  Details can be found in Tab.~\ref{tab:details},
where the trajectory numbers shown are those trajectories where
measurements are done.  The length of each trajectory is $0.5$
molecular dynamics time units, and measurements are done every 10
trajectories.  The jackknife re-sampling method is used throughout
our analysis and all the errors we present in this paper are only
statistical errors.

\begin{table}
  \centering
  \begin{tabular}{llcclllll}
    $\beta$ & Size & $m_q$ & Trajectories
    & $\pbp$ & $m_\rho$ & $r_0$ & $a^{-1}/\mathrm{GeV}$ \\
    \hline\hline
    0.58
    &  $16^3\times32$  &  0.025  &  $1330\sim2760$  &   0.09973(27)  &   0.812(11)  &   4.39(56)  &  1.73(22)  \\
    &                  &  0.015  &  $880\sim1950$   &   0.06582(13)  &   0.619(13)  &   5.05(78)  &  1.99(31)  \\
    &  $24^3\times32$  &  0.025  &  $1060\sim3390$  &  0.100381(67)  &  0.7832(30)  &  4.126(96)  &  1.628(38) \\
    &                  &  0.015  &  $960\sim2930$   &   0.06652(11)  &  0.6126(28)  &   5.10(11)  &  2.014(45) \\
    \hline
    0.56
    &  $16^3\times32$  &  0.024  &  $970\sim4920$   &   0.13643(20)  &  0.9431(38)  &   3.19(18)  &  1.259(70) \\
    &                  &  0.016  &  $1040\sim3730$  &   0.10147(26)  &   0.803(12)  &   3.68(15)  &  1.451(58) \\
    &  $24^3\times32$  &  0.024  &  $1010\sim3340$  &   0.13668(14)  &  0.9693(69)  &  3.120(48)  &  1.231(19) \\
    &                  &  0.016  &  $1040\sim3190$  &   0.10208(12)  &  0.8085(93)  &  3.793(97)  &  1.497(38) \\
    &                  &  0.008  &  $1000\sim2970$  &   0.06148(16)  &  0.6022(73)  &  4.716(92)  &  1.861(36) \\
    \hline
    0.54
    &  $16^3\times32$  &  0.03   &  $1010\sim6220$  &   0.23100(20)  &   1.258(17)  &  2.197(52)  &  0.867(21) \\
    &                  &  0.02   &  $990\sim5300$   &   0.19646(28)  &   1.176(19)  &  2.350(47)  &  0.927(19) \\
    &                  &  0.01   &  $1030\sim5520$  &   0.14464(37)  &   0.993(14)  &  2.849(51)  &  1.124(20) \\
    &  $24^3\times32$  &  0.01   &  $1070\sim2860$  &   0.14393(39)  &   1.022(17)  &  2.830(48)  &  1.117(19) \\
  \end{tabular}
  \caption{A brief summary of our simulation parameters and results.
  The value of $a^{-1}$ is determined from measuring $r_0$ and assuming
  $r_0$ for $N_f = 8$ has a physical value of $0.5\mathrm{fm}$.}
  \label{tab:details}
\end{table}

\begin{figure}
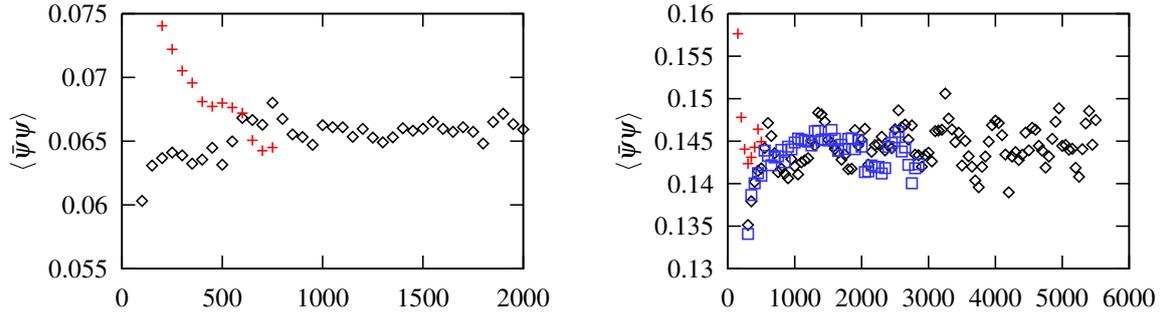

  \centering
  \includegraphics[width=.47\textwidth]{fig/pbp_8m015_b058.mps}
  \qquad
  \includegraphics[width=.47\textwidth]{fig/pbp_8m01_b054.mps}
  \caption{Evolution of $\pbp$.  Left: $\beta=0.58$, $m_q=0.015$.
    Right: $\beta=0.54$, $m_q=0.01$.  In both plots, the data is binned
    in blocks of 50 trajectories.  Black diamonds and red crosses
    represents ordered and disordered starts, respectively, with a
    lattice size of $16^3\times32$.  Blue squares in the right plot
    are from an ordered start with a lattice size of $24^3\times32$.}
  \label{fig:pbp_evolve}
\end{figure}

For all three $\beta$ values, we start our RHMC simulations from
both ordered and disordered configurations.  In Fig.~\ref{fig:pbp_evolve}
we show the evolutions of $\pbp$ for $\beta=0.58$ and $\beta=0.54$.
We see that ordered and disordered starts have values of $\pbp$
which agree after thermalization.  The metastability for eight
flavor simulations with the Wilson gauge action seen in \cite{Brown:1992fz}
does not appear in our current work.  In addition to the change in
action, the earlier work used the inexact R algorithm, while here
we use the exact RHMC.  While we have not fully investigated what is
responsible for the disappearance of metastability, it is very helpful
to see that it is not present in the current simulations.

\begin{figure}
  \centering
  \includegraphics[width=.6\textwidth]{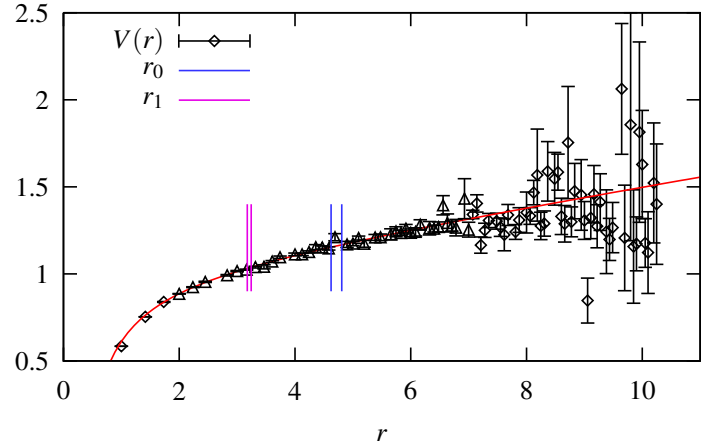}
  \caption{Heavy quark potential measured on ensemble of
    $\beta=0.56$, $m_q=0.008$, with lattice size of $24^3\times32$.}
  \label{fig:hqpot}
\end{figure}

\begin{figure}
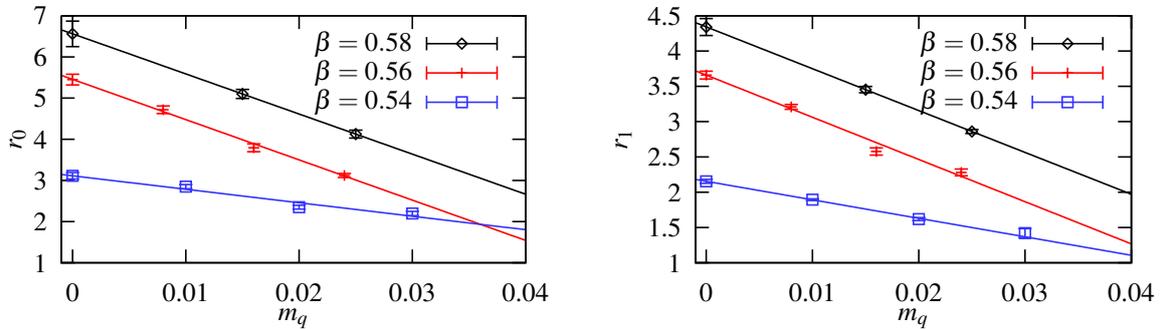

  \centering
  \includegraphics[width=.47\textwidth]{fig/r0_ext.mps}
  \qquad
  \includegraphics[width=.47\textwidth]{fig/r1_ext.mps}
  \caption{Extrapolation to the chiral limit of $r_0$ and $r_1$.}
  \label{fig:r0_r1_ext}
\end{figure}

The heavy quark potential is measured on our ensembles using the method
in \cite{Li:2006gra}, and we fit the potential to the form
\begin{equation}
  \label{eq:1}
  V(r) = V_0 - \frac{\alpha}{r} + \sigma r .
\end{equation}
Fig.~\ref{fig:hqpot} shows the heavy quark potential for the ensemble
with $\beta=0.56$, $m_q=0.008$ on a $24^3\times32$ lattice.  The red
curve is the fit to Eq.~(\ref{eq:1}) over the range of data points
denoted by triangle symbols, and the data points shown as diamond
symbols are left out of the fit.  We can clearly see confining
behavior from the shape of the potential.  Values of $r_0$ and $r_1$
are obtained from the fit and we extrapolate them linearly to the zero
quark mass limit as shown in Fig.~\ref{fig:r0_r1_ext}.

\begin{figure}
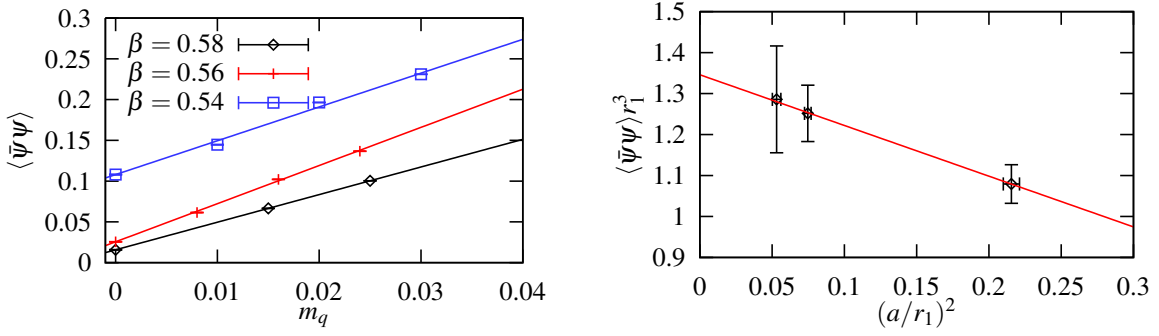

  \centering
  \includegraphics[width=.47\textwidth]{fig/pbp_ext.mps}
  \qquad
  \includegraphics[width=.47\textwidth]{fig/pbp_limit.mps}
  \caption{The chiral limit extrapolation of the chiral condensate (left
  panel), and the continuum limit of the chiral limit values.
    $\pbp$ is not renormalized.}
  \label{fig:pbp_ext}
\end{figure}

It is important to check whether $N_f = 8$ QCD has dynamical chiral symmetry
breaking, which gives $\pbp$ a non-zero value in the zero quark
mass limit.  In Fig.~\ref{fig:pbp_ext}, we show linear extrapolations
of $\pbp$ to $m_q = 0$ (left panel) and the scaling
behavior of $\pbp(m_q=0)$ versus $a^2$ (right
panel).  The data clearly support a
non-zero chiral condensate in the continuum limit.

\begin{figure}
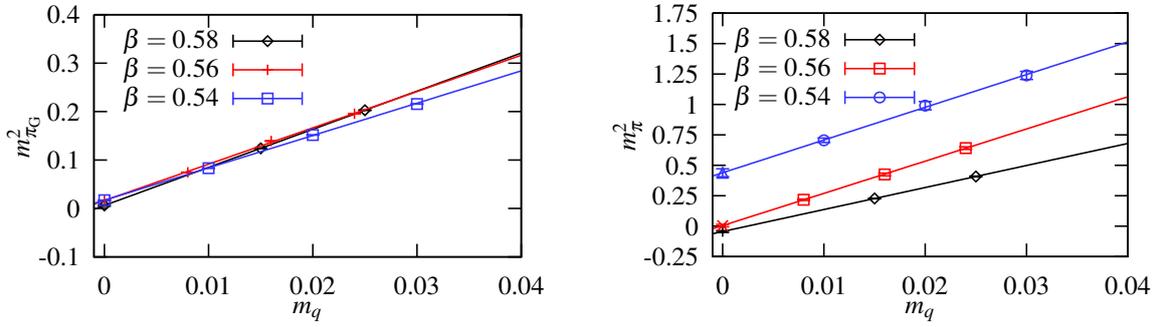

  \centering
  \includegraphics[width=.47\textwidth]{fig/pion_ext.mps}
  \qquad
  \includegraphics[width=.47\textwidth]{fig/pion2_ext.mps}
  \caption{Chiral extrapolation of pion mass.  Left: Goldstone Pion
    mass.  Right: Non-Goldstone Pion mass -- Scalar channel which
    corresponds to $r^{\sigma_s\sigma_{123}}=1++$, of the meson
    propagators.}
  \label{fig:pion_ext}
\end{figure}

If $N_f = 8 $ QCD is in a phase with broken chiral symmetry, there
should be a pseudo-Goldstone boson whose mass vanishes in the chiral
limit.  Since staggered fermions have a remnant chiral symmetry,
we should be able to see this Goldstone particle in our simulations.
We have measured the masses for both the Goldstone pion ($\pi$) and
the local non-Goldstone pion ($\pi_2$) in our simulations.  In
Fig.~\ref{fig:pion_ext}, the left panel shows a linear extrapolation
of $m_\pi^2$ versus $m_q$.  (Clearly there are chiral logarithm
corrections to a simple linear extrapolation, but the coefficient
of the chiral log term in the continuum goes as $1/N_f$, which may
make the effects harder to see here than in physical QCD.   There
are also taste breaking effects in the chiral limit.  We have not
attempted to include any NLO chiral terms of this kind.) The
extrapolated Goldstone mass is consistent with zero, supporting the
conclusion that $N_f =8$ QCD is in the chirally broken phase.  The
right panel of Fig.~\ref{fig:pion_ext} shows the extrapolation of
$m_{\pi_2}^2$ to the chiral limit, which is clearly non-zero for
our coarsest lattice.  For the finer lattices, the taste-symmetry
breaking is much smaller.

We have done a linear extrapolation of our measured values for $m_\rho$
to the chiral limit, and then extrapolated these to the continuum limit,
where we find a non-zero value for $m_\rho(m_q = 0)$.

%\begin{figure}
%  \centering
%  \includegraphics[width=.47\textwidth]{fig/rho_ext.mps}
%  \qquad
%  \includegraphics[width=.47\textwidth]{fig/rho_limit.mps}
%  \caption{Chiral extrapolations of the rho mass, and the $a^2$ dependence
%  of $m_{\rho}(m_q = 0)$.}
%  \label{fig:rho_ext}
%\end{figure}

\section{Conclusions and outlook}
\label{sec:conclusions-outlook}

We have shown data from lattice simulations of QCD with 8 degenerate
quark flavors that supports, in the chiral and continuum limits,
a non-zero value for $r_0$ and $r_1$ from the heavy quark potential,
a non-zero value for $\pbp$, a massless $\pi$ and a massive $\rho$.
This argues that the conformal phase for QCD with $N_f$ fermions
in the fundamental representation must begin with $N_f > 8$.

%We have used simple linear fits for the chiral extrapolations of
%$r_0$, $r_1$, $\pbp$, $m_\rho$ and $m_\pi^2$ and some small deviations
%are visible in our data.  Such effects do not appear large enough to
%alter our conclusions.  Additionally, we have done simulations on two
%lattice volumes and do not see any large effects as the volume is changed.

Having presented arguments for the behavior of 8 flavor QCD in the chiral
and continuum limits, we are considering looking for evidence of a zero
temperature conformal phase in low energy QCD observables with more flavors.

\section*{Acknowledgments}

We are thankful to all members of the RBC collaboration for their
helpful discussions.  Our evolution and propagator measurement code
is from Michael Cheng, with minor modifications, and the heavy quark
potential measurement code is from Min Li.  Our calculations were
done on the QCDOC at Columbia University and NY Blue at BNL.  This
research utilized resources at the New York Center for Computational
Sciences at Stony Brook University/Brookhaven National Laboratory
which is supported by the U.S. Department, of Energy under Contract
No. DE-AC02-98CH10886 and by the, State of New York.

\bibliographystyle{JHEP-2}
\bibliography{ref}

\end{document}